\documentclass[%
 reprint,
 superscriptaddress,
 showpacs,amsmath,amssymb,
 aps,
floatfix,
nofootinbib,
twocolumn
]{revtex4-2}
\usepackage{url}
\usepackage[breaklinks]{hyperref}
\PassOptionsToPackage{hyphens}{url}\usepackage{hyperref}
\usepackage{mathpazo}
\usepackage{times}
\usepackage{array}
\usepackage{siunitx}
\usepackage{dcolumn}%
\usepackage{footnote}
\usepackage{makecell,multirow}
\usepackage{booktabs}
\usepackage[utf8]{inputenc}
\usepackage{tikz}
\usetikzlibrary{shapes,arrows}
\usepackage{multirow}
\usepackage{siunitx}
\usepackage{amsmath}
\usepackage{bm}% bold math
\usepackage[mathlines]{lineno}
\usepackage{braket}
\usepackage{color,soul}
\usepackage[justification=justified]{caption}
\usetikzlibrary{shapes.geometric}
\usepackage{changepage}
\usepackage[bottom]{footmisc}
\usepackage{float}
\usepackage[american,siunitx]{circuitikz}
\usetikzlibrary{arrows,calc,positioning}
\usepackage{comment}
\usepackage{gensymb} %for the degree symbol
\usepackage{subfig}
\usepackage{graphicx}% Include figure files
\newsavebox{\foobox}

\tikzset{ar/.style={-latex,shorten >=-1pt, shorten <=-1pt}}
\begin{document}
\preprint{APS/123-QED}

\title{Antenna-coupled microstrip summing network for wave-like dark matter searches}

\author{Chelsea Bartram}
\affiliation{SLAC National Accelerator Laboratory, 2575 Sand Hill Road, Menlo Park, California 94025, USA}

\author{Taj A. Dyson}
\affiliation{Department of Physics, Stanford University, 382 Via Pueblo Mall, Stanford, CA, 94305, USA}

\author{Jacob Laurel}
\email[Corresponding author: ]{laurelj1@uci.edu}
\affiliation{SLAC National Accelerator Laboratory, 2575 Sand Hill Road, Menlo Park, California 94025, USA}
\affiliation{University of California Irinve, Irvine, California, 92617, USA}

\author{Sephora Ruppert}
\affiliation{Department of Physics, Stanford University, 382 Via Pueblo Mall, Stanford, CA, 94305, USA}

\author{Matthew O. Withers}
\affiliation{Department of Physics, Stanford University, 382 Via Pueblo Mall, Stanford, CA, 94305, USA}

\author{Osmond Wen}
\affiliation{Department of Physics, Stanford University, 382 Via Pueblo Mall, Stanford, CA, 94305, USA}

\author{Andrew K. Yi}
\email[Corresponding author: ]{andrewyi@slac.stanford.edu}
\affiliation{SLAC National Accelerator Laboratory, 2575 Sand Hill Road, Menlo Park, California 94025, USA}

\author{Barkotel F. Zemenu}
\affiliation{Department of Physics, Stanford University, 382 Via Pueblo Mall, Stanford, CA, 94305, USA}

\author{Chao-Lin Kuo}
\affiliation{SLAC National Accelerator Laboratory, 2575 Sand Hill Road, Menlo Park, California 94025, USA}
\affiliation{Department of Physics, Stanford University, 382 Via Pueblo Mall, Stanford, CA, 94305, USA}

\date{\today}

%\linenumbers\relax % Commence numbering lines

\begin{abstract}
A commonly used technology for axion and other wave-like dark matter searches is a microwave resonator. In such experiments, the detected signal strength depends on the coupling between the resonator and the microwave transmission line that delivers power to the amplifier.
In Axion Dark Matter eXperiment (ADMX) and other cavity searches, this coupling is typically achieved using a compact wire antenna inserted into the cavity.  While this approach is adequate for right-cylindrical cavities, it can become difficult to efficiently extract power from large-volume, multi-wavelength ($V \gg \lambda^3$) resonators targeted for higher frequencies, such as those designed by ADMX-VERA (Volume Enhanced Resonant Axion). % that have cavities with two extended dimensions.
Here, we present a solution based on a microwave slot-antenna array capacitively coupled to a microstrip ``summing network,” which coherently combines signals from multiple transmission lines into a single output. This summing architecture is adapted from designs originally developed for cosmic microwave background (CMB) detectors. We demonstrate the design and characterization of a new high-frequency resonator that employs this readout scheme in the $19-21$ GHz range. Finally we discuss the potential to reach Dine–Fischler–Srednicki–Zhitnitsky (DFSZ) sensitivity using multiple summing network cavities fitted into a large volume.
\end{abstract}
\maketitle

\section{Introduction}

Axions are among the most theoretically well-motivated dark matter (DM) candidates~\cite{ABBOTT1983133,PRESKILL1983,DINE1983}. In addition to being capable of accounting for the full DM abundance, they naturally arise as a solution to the strong charge-parity (CP) problem via the Peccei–Quinn mechanism~\cite{Peccei1977Sept, weinberg,wilczek}. A variety of axion models have been developed within this framework, with two benchmark examples being the Kim–Shifman–Vainshtein–Zakharov (KSVZ)~\cite{KIM1979,SHIFMAN1980} and Dine–Fischler–Srednicki–Zhitnitsky (DFSZ)~\cite{DINE1981,Zhitnitsky:1980tq} models.
These models generically predict very weak couplings between the axion and electromagnetic fields, requiring highly sensitive experiments operating in cryogenic environments for detection. Moreover, the viable axion mass range spans many orders of magnitude, making it essential for detectors to scan across a wide frequency range while maintaining high sensitivity.

For experiments such as ADMX~\cite{admx-recent}, the axion haloscope~\cite{SikivieHaloscope} satisfies these requirements using a high quality factor ($Q$) resonant cavity immersed in a strong magnetic field. In this setup, axions are converted into photons with frequencies proportional to the axion mass, producing a resonantly enhanced signal within the cavity that can be detected.

An important figure of merit for any haloscope is the scan rate~\cite{SikivieScanRate, HagmannScanRate}, defined as the rate at which the cavity can be tuned while achieving a target signal-to-noise ratio (SNR) for a benchmark axion signal. This quantity is given by

\begin{equation}
\label{scanRate}
\begin{split}
        \frac{\mathrm{d}\nu}{\mathrm{d} t} & =  543 \frac{\text{MHz}}{\text{yr}} \left(\frac{\beta}{1+\beta}\right)^2 \left(\frac{g_\gamma}{0.36}\right)^4 \left(\frac{\nu}{740\text{ MHz}}\right)^2 \\ & \times\left(\frac{\rho}{0.45 \text{ GeV / cm}^3}\right)^2\left(\frac{3.5}{\text{SNR}}\right)^2\left(\frac{B}{7.6\text{ T}}\right)^4 \\& \times \left(\frac{V}{136 \text{ l}}\right)^2 \left(\frac{Q_L}{30000}\right)\left(\frac{C}{0.4}\right)^2\left(\frac{0.2\text{ K}}{T_{\text{sys}}}\right)^2,
\end{split}
\end{equation}
where $g_\gamma$ is a model dependent parameter ($0.36$ for DFSZ and $-0.97$ for KSVZ), $\nu$ is the resonant frequency of the cavity, $\rho$ is the background DM density, $B$ is the magnetic field strength, $V$ is the volume of the cavity, and $T_{\text{sys}}$ is the temperature of the system (for ADMX, that is a few $100$ mK~\cite{ADMX_Noise_PRD}). The parameter $\beta$ is the coupling of the read-out system with the cavity, which is typically a bare wire inserted into the cavity, and expresses how much power is being removed from the cavity by the read-out system. $Q_L$ is the loaded quality factor. $Q_{L}$ is related to the unloaded quality factor of the cavity $Q_{0}$ through $\beta$:
\begin{equation}
Q_{0} = (1+\beta)Q_{L}.
\end{equation}
Thus the scan rate dependence for on $\beta$ is 
\begin{equation} \label{eq_scan_rate_beta}
    \mathrm{d}\nu / \mathrm{d} t \propto \beta^{2}/(1 + \beta)^{3} .
\end{equation}
 In this work, $\beta$ will be the main parameter in consideration of maximizing scan rate.

$C$ is the form factor, which quantifies the alignment of the resulting photon's electric field $\vec{E}$ and the external magnetic field $\vec{B}$, given by
\begin{equation}
    C=\frac{\left| \int  \vec{E} \cdot \vec{B} dV\right|^2}{B^2V\int |\vec{E}|^2dV}.
\end{equation}

One well motivated frequency range is for post-inflationary axions~\cite{marsh2016axion, borsanyi2016axionqcd, klaer2017axionmass, graham2018stochastic, takahasi2018qcdaxion, buschmann2022axionmass} which corresponds to higher frequencies around 4 to 30 GHz. As ADMX continues to scan up into higher GHz frequencies and enter this post-inflationary regime, it becomes increasingly important to maintain or improve scanning rate performance. If one continues to use scaled cylindrical cavities such as those used in ADMX, the volume goes as
\begin{equation}
    V \propto \nu^{-3},
\end{equation}
so based on eq.~(1), the scan rate scales as
\begin{equation}
    V^2\cdot\nu^2\propto \nu^{-4}.
\end{equation}
Including additional frequency dependencies from the quality factor, $(Q_L\propto\nu^{-\frac{2}{3}})$, and the system noise $(T_{\mathrm{sys}}\propto\nu^{-1})$, the overall scaling becomes
\begin{equation}
    \frac{\mathrm{d}\nu}{\mathrm{d}t}\propto \nu^{-\frac{14}{3}}.
\end{equation}

This strong frequency dependence presents a key challenge: as the search moves to higher frequencies, the scanning rate decreases rapidly. The ADMX-VERA (Volume-Enhanced-Resonant-Axion) program addresses this limitation through new cavity geometries designed to maintain large effective volume at high frequencies. The main goal of this work is to solve a potential drawback of those cavities: maintaining sufficiently high antenna coupling $\beta$ to optimize scan rate. 

\section{Coupling to ADMX-VERA Cavities} \label{ADMX Overview}
Over the last few years, the ADMX-VERA program has developed multiple high-volume resonators. The first broad class of geometries derive from variations of a rectangular cavity with one narrow dimension and two extended dimensions~\cite{Kuo_2020NovelCavities}. For the wavelength $\lambda$ of a target electromagnetic wave, the narrow dimension would be about half that wavelength ($\sim \lambda/2$). This class is exemplified by the single wedge cavity~\cite{singleWedge_Taj2024}, as well as its three-fold extension, the triple wedge cavity~\cite{maximizing}. A second class of designs utilizes a ``stiff" global eigenmode arising in strongly coupled multi-cell structures, as realized in the beehive cavity~\cite{withers2025beehive}. 

This paper primarily focuses on coupling to the first class of resonators: thin-shell cavities.  The axion-sensitive eigenmodes in these structures are analogous to the TM$_{010}$ mode of a cylindrical cavity. In these modes, the Poynting vector is predominantly oriented along the narrow dimension of the cavity. The resonance can therefore be viewed as a standing plane wave confined between the two (nearly) parallel plates that define this dimension.  Tuning can be achieved by a lateral displacement of one set of walls or by adopting a conical geometry~\cite{Kuo_2021}.

In principle, the two extended dimensions can be made arbitrarily large, leading to a cavity with $V \gg \lambda^3$. One of these extended dimensions can be wrapped into a space-filling curve, allowing the cavity to be accommodated within a solenoid bore~\cite{Kuo_2020NovelCavities, Kuo_2021}. However, before declaring the volume problem solved, one must ensure that the energy stored in such a cavity can be efficiently extracted.

Traditional wire probes are no longer suitable in this regime. For example, in the aforementioned prototype wedge cavity, a coupling of only $\sim 0.1$ was achieved using a wire probe, whereas axion haloscopes require an antenna coupling of $\beta = 2$ for optimal performance~\cite{haystac_analysis}. Additionally, the electric field distribution within the cavity can be affected by perturbations of the cavity boundary conditions. This sensitivity can result in localized modes, concentrated into a small area of the overall volume. 

\begin{figure}
% \centering
    \includegraphics[width=\linewidth]{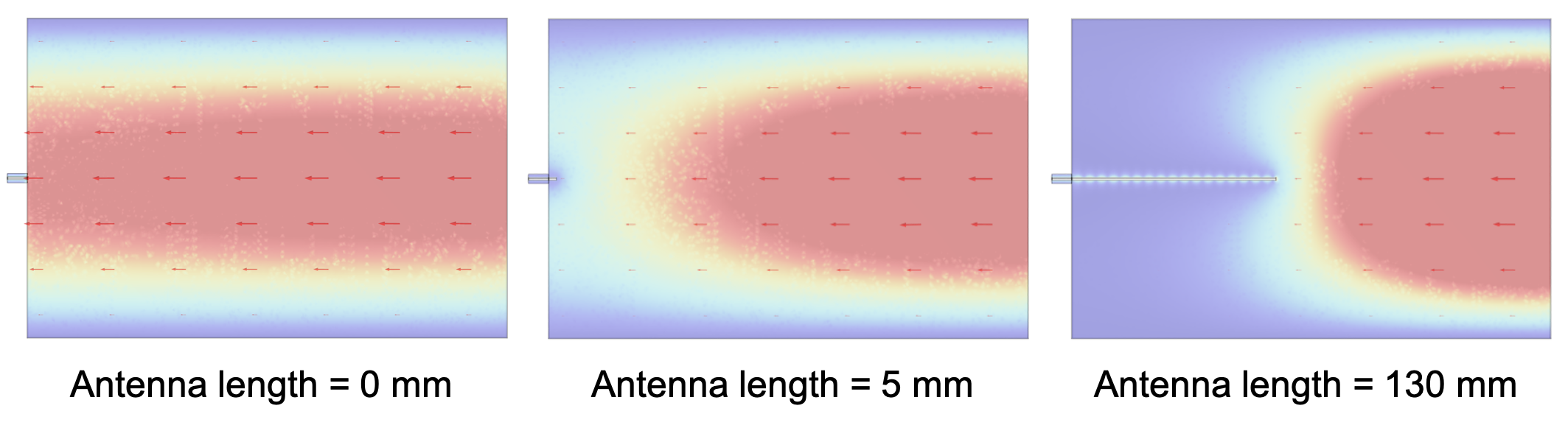}
    \caption{A simulated electric field perturbation of a wire antenna inside a rectangular cavity for a TM${}_{110}$ mode. Without any antenna length inside the cavity (left, 0 mm), the electric field distribution is preserved; however, as the antenna is inside the cavity (center, 5 mm, and right, 130 mm) the electric field distribution begins to change inside the cavity. The arrows show the orientation of the electric field, while the colors show its strength, with red being stronger.
    }
    \label{fig_ant_perturb}
\end{figure}

Even the insertion of a wire probe can disrupt the uniformity of the electric field within the cavity. As the volume increases, attempts to maintain $\beta$ by increasing the strength of the coupling often lead to mode localization, resulting in a lower form factor~\cite{maximizing}. This effect can be seen in various cavity geometries, and here we focus on a rectangular cavity which will be the baseline design. Fig.~\ref{fig_ant_perturb} illustrates the change in electric field distribution as a wire antenna is inserted inside a rectangular cavity in an attempt to couple with the TM${}_{110}$ mode. 

An alternative perspective is that the axion-sensitive mode in VERA cavities oscillates coherently over a volume $V \gg \lambda^3$. To function as an axion haloscope, the energy stored in this spatially distributed mode must be coherently directed onto a microwave transmission line with adequate total coupling. 
To meet these requirements, we take inspiration from cosmic microwave background (CMB) detectors that employ distributed slot-antenna arrays to couple to the spatial modes provided by the telescope optics~\cite{CMB_slotArray}. In this approach, coherent signals collected by the slot antennas are capacitively coupled to a network of microstrip transmission lines, which we refer to as the microstrip summing network.

This summing network addresses the coherence problem by coupling to multiple locations within the cavity with equal electrical distances, effectively integrating signals across the entire mode volume.  In contrast, a conventional compact antenna typically couples only to a region of size $\lambda^3$. The slot antennas perturb the spatial distribution of the axion-sensitive eigenmode weakly, thereby maintaining a higher form factor. Moreover, the summing network acts as a mode-selective device, since signals from higher-order modes interfere destructively during summing~\cite{Kuo_2020NovelCavities}.

\section{A rectangular cavity with a microstrip summing network}

As a case study, we consider a $\lambda/2$-thick rectangular slab cavity coupled to a microwave summing network. As discussed earlier, this parallel-plate cavity geometry serves as a useful starting point for a broad class of VERA designs.  We will begin with a discussion of a summing network fabricated on a printed circuit board (PCB) using slot antennas and a microstrip summing path that can be used as a readout chain. After this, we compare a rectangular cavity using a single wire antenna readout and the summing network readout using finite-element analysis simulations. In the following section, we demonstrate the performance of this design via benchtop measurements and compare the results with simulation.

\subsection{Cavity and its eigenmodes} \label{subsec_cavity}

The basis of the cavity design is a rectangular cavity with a height that can vary, and the other two dimensions fixed at 203.2 mm $\times$ 304.8 mm. The rectangular microwave cavity has several resonant modes. Out of those, the TM${}_{110}$ mode is of primary interest as it has the highest form factor, and therefore maximizes the axion signal power from the cavity assuming a relatively constant, static external magnetic field in a single direction. This mode also has the lowest resonant frequency among the TM modes. This work focuses on a frequency range centered on 20 GHz, which corresponds to a resonator height of approximately 7.5 mm.  The full volume of the cavity is 138 $\lambda^3$ at the center frequency. 

\begin{figure}
    \centering
    \includegraphics[width=0.8\linewidth]{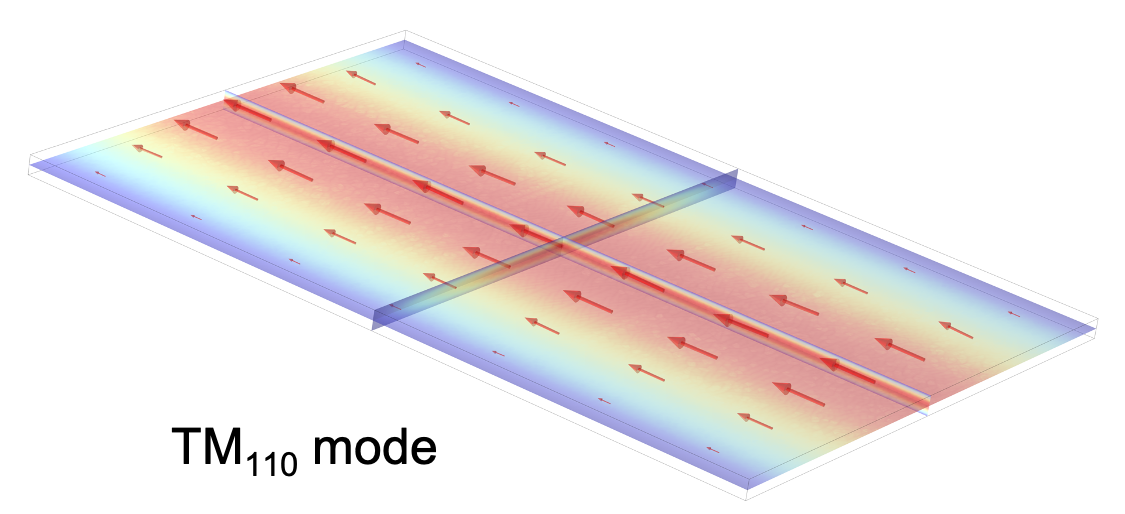} \\
    \includegraphics[width=0.8\linewidth]{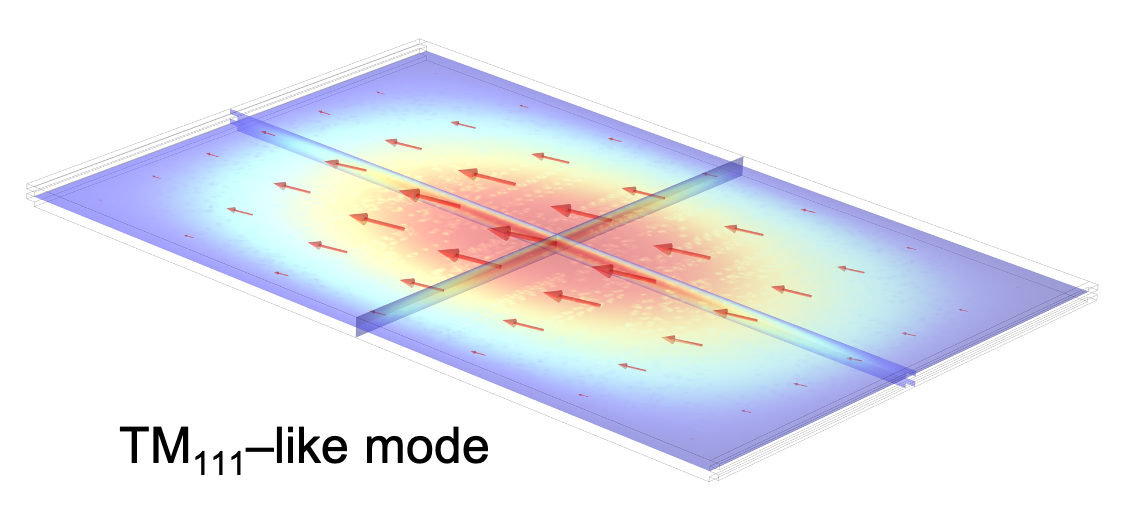}
    \caption{Electric field distribution for lowest order mode of the cavity without (top, TM${}_{110}$ mode) and with RF chokes (bottom, TM${}_{111}$-like mode) simulated with COMSOL (not including the summing network). Both resonant modes were found in eigenfrequency mode, around 20 GHz where the top and bottom plates were 7.5 mm apart. The TM${}_{111}$-like mode has a resonant frequency that is about 6 MHz higher. The arrows show the orientation of the electric field, while the colors show its strength, with red being stronger.}
    \label{COMSOL_modes}
\end{figure}

The walls of the cavity are made of Al-6061 plates, except for the bottom wall, which is formed by the copper ground plane of the PCB that contains the slot-antenna array and summing network. The two sidewalls that are perpendicular to the electric field inside the cavity were modified to have corrugations, called radio frequency (RF) chokes, to modify the boundary conditions on the electric field. This changes the resonance structure of the TM${}_{110}$ mode to look closer to a TM${}_{111}$-like mode. Both mode structures, simulated with COMSOL, are shown in Fig.~\ref{COMSOL_modes}. For both cases, the resonance frequency mainly depends on the distance between the top and bottom plates ($z$-axis). 

\subsection{Design of the summing network}
\label{ST_design}

\begin{figure}
    \centering
    \includegraphics[width=0.9\linewidth]{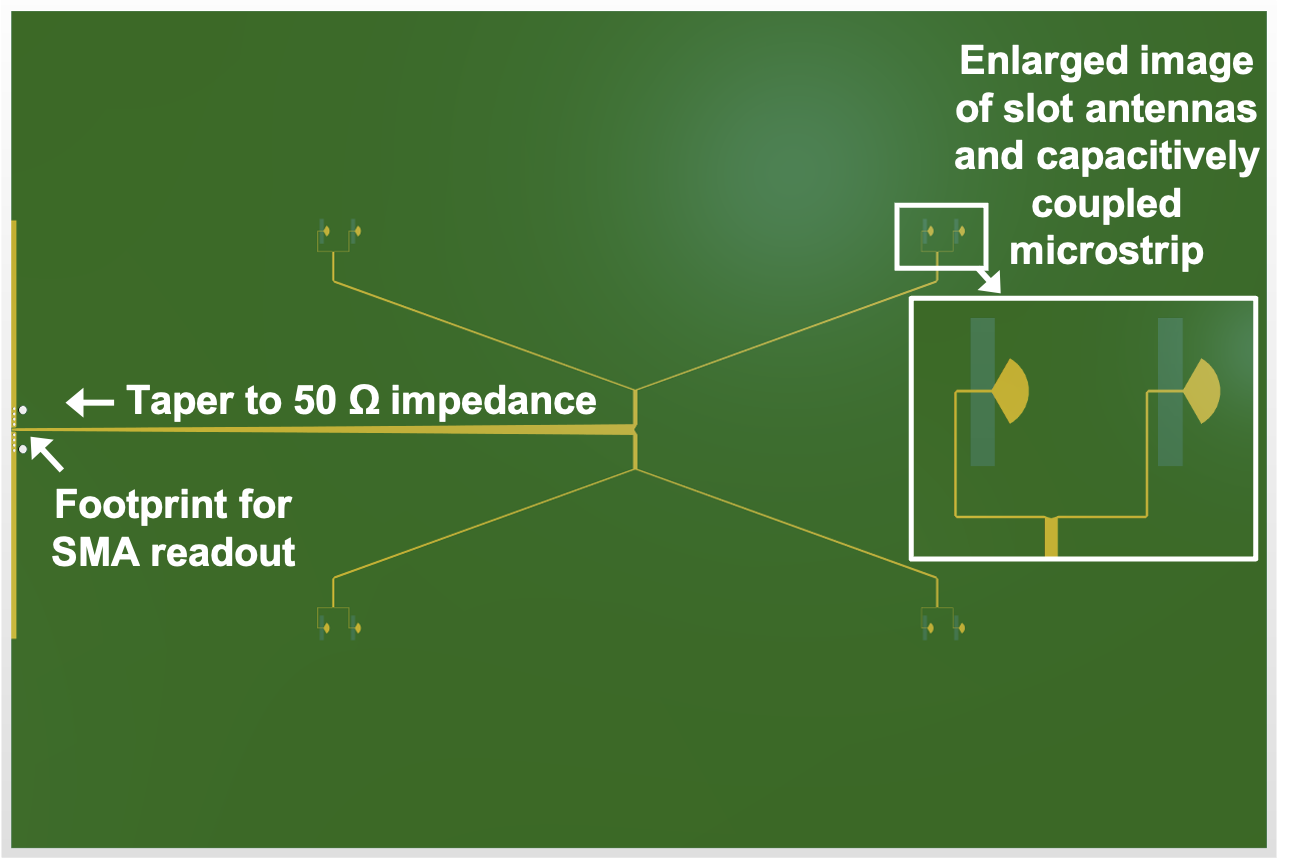}
    \caption{Schematic of the 8-slot microstrip summing network on a PCB. The other side not shown in the figure is a copper plate with rectangular cutouts, shown as faint rectangles in the inset, that act as slot antennas.}
    \label{ST schematic}
\end{figure}

The PCB design of the summing network contains many different components of interest, shown in Fig.~\ref{ST schematic}. The outline of the PCB is the same as the rectangular slab, being 203.2 mm $\times$ 304.8 mm, with one side, the Cu ground plane, acting as one of the cavity walls.

The Cu ground plane of the PCB contains an array of slot antennas to read out the resonances in the cavity. For this prototype, the antenna array consists of eight slot antennas arranged symmetrically along the PCB as shown in Fig.~\ref{ST schematic}. The slots are elongated cutouts in the otherwise continuous ground plane. Each slot antenna is a rectangle with dimensions 1 mm $\times$ 6 mm. The electric field of the target mode, TM${}_{110}$ mode, is parallel to the longest cavity dimension, while the antennas are aligned such that the electric field is parallel to the shorter dimensions of the slots. The singly polarized, axion-sensitive resonant mode induces a current on the ground plane and thereby a voltage across these slots. Slot antennas are designed to be broadband, and for the dimensions given the bandwidth is roughly a 20\% frequency range (about 4 GHz) centered at 20 GHz. % Show bandwidth of antenna? 

Beneath this Cu layer lies a 0.254 mm thick layer of Rogers 5880 substrate. This substrate has a  dielectric constant of 2.2 and is selected for its very low loss tangent. The substrate also affects the width of the microstrip depending on the impedance it needs to match.

The Au-plated Cu microstrip network is patterned on top the Rogers 5880 layer, opposing the slotted ground plane. At each branch of the network a capacitive fan is connected to the microstrip that runs across the slot on the other side of the Cu plane. These capacitive fans have a reactive impedance that cancels out the reactive impedance of each corresponding slot antenna. All fan capacitors are oriented in the same direction to coherently capture the singly polarized mode produced by axion conversion in a uniform external magnetic field. This singly polarized, axion-sensitive mode induces a current on the ground plane and thereby produces voltage across slots. The signal is then carried through the various branches of the microstrip.

The microwave summing netowrk, or ``summing tree" design closely follows that of the BICEP CMB detectors, described in~\cite{CMB_slotArray}.  Each microstrip begins with a width of 4 mils (0.1016 mm), close to a typical minimal feature size in commercial PCBs. This width, combined with the properties of the Rogers 5880 substrate, provides an impedance of 125 $\Omega$. When the microstrip is bent at an angle to meet with another ``branch" of the summing network, mitered bends are designed to reduce the loss from reflection. Symmetry in the slot configuration ensures equal electrical path lengths from each slot to the end of the microstrip, allowing signals from all antennas to add coherently. At each junction where the microstrip combines multiple antenna feeds, the trace width increases to maintain impedance matching among all lines. The number of slot antennas that each composite branch carries acts as a weight factor that is then added in the same way parallel resistors combine resistance. In the case of the eight-slot array, all weights are same at each junction, so the width of the microstrip doubles, and the impedance of the microstrip is halved. As such, the combined impedance at the combined junction of microstrip branches is $Z_{1} = 15.6$ $\Omega$. Finally, each microstrip is tapered to gradually transition to a 50 $\ohm$ impedance. An exponential taper, which has been approximated to a piecewise linear taper in design, is applied. The impedance of the given taper is
\begin{equation}
    Z(z) = Z_{0}e^{-\alpha z}, 
\end{equation}
where $Z_{0} = 50$ $\ohm$ and $\alpha = \frac{1}{z_{0}} \ln({Z_{1}}/{Z_{0}})$. The $z_{0}$ in the calculation for $\alpha$ is the length of the taper, equal to roughly 149.4 mm, or about half the length of the PCB for this work. 

At the end of the microstrip, the signal transitions to a coplanar waveguide (CPW) section, with vias through the dielectric connecting the CPW ground strips to the underlying copper plane. The CPW signal line terminates at an edge-mounted SMA connector rated for operation up to 26 GHz.

\subsection{Integration of microwave cavity}
Since the microstrip summing network is being used as a readout system for an axion search, it is attached to the rectangular cavity described in Sec~\ref{subsec_cavity} which provides the volume and resonant mode for axion-to-photon conversion. The longer side of the rectangular cavity aligns with the direction of slot antenna excitation. The resonant mode is coupled to the antennas, and read out after being transferred and combined through the entirety of the microstrip, as discussed in Sec.~\ref{ST_design}.

\begin{figure} [h]
    \centering
    \includegraphics[width=0.9\linewidth]{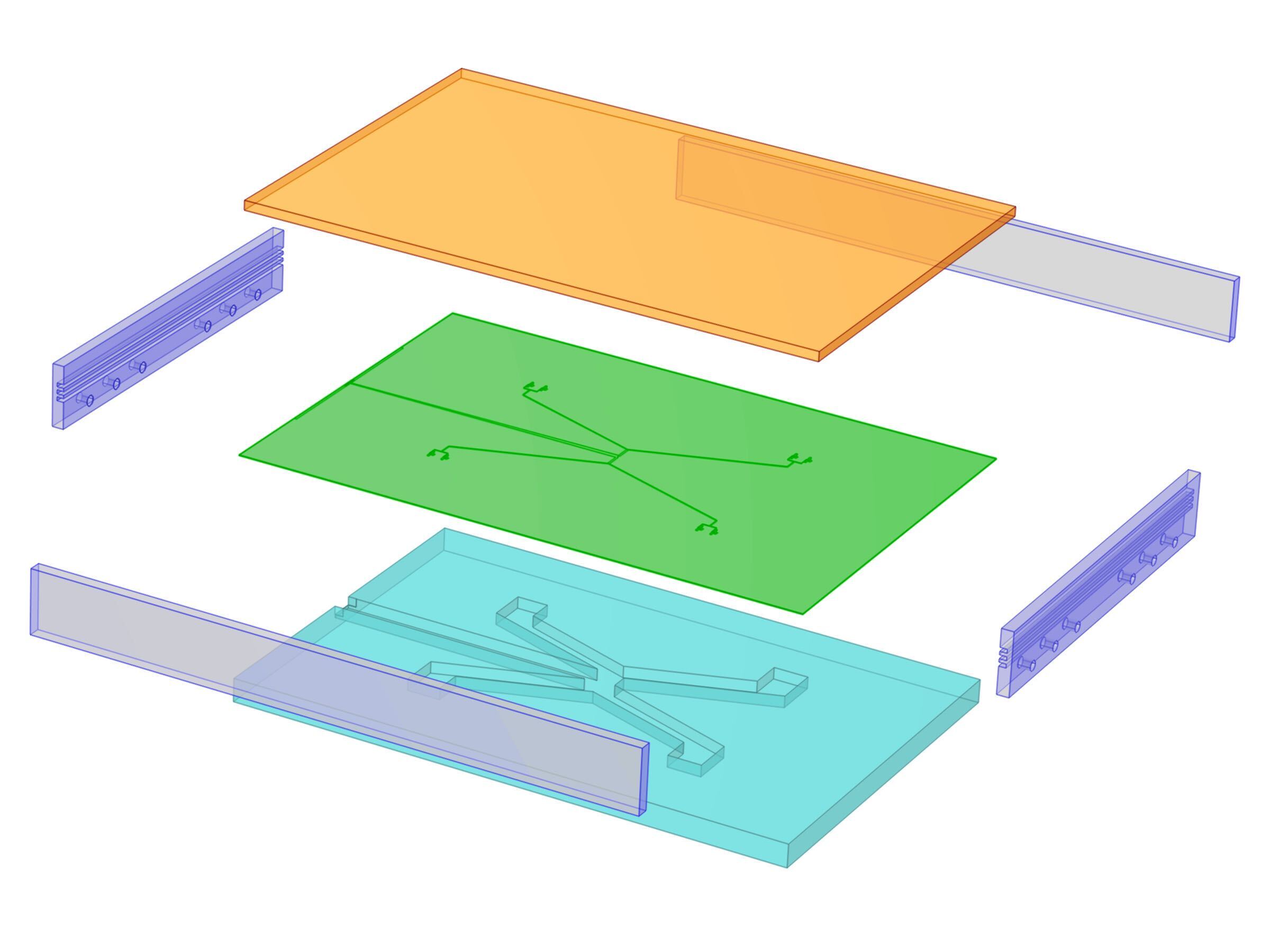}
    \caption{An exploded view of the components of the cavity coupled to a microstrip summing network. At the center, the summing network PCB (green) is attached to a bottom plate with trenches (light blue). All sides of the rectangular cavity are walled (shown in dark blue), and the left and right walls have corrugations. A top plate (orange) moves to adjust the height of the cavity. (In the simulation, the front and back walls are omitted since they can be set with simpler boundary conditions.) 
    }
    \label{fig_comsol_cavity}
\end{figure}

A set of COMSOL simulations was conducted to compare the improved antenna coupling of the summing network to that of a conventional single wire antenna. Fig.~\ref{fig_comsol_cavity} shows the design of the cavity rendered in COMSOL. The top and bottom plates have a variable height and for this work we focus on the distances between 7.1 and 7.9 mm, corresponding to a frequency range of $19-21$ GHz for the lowest order mode. The cavity also utilizes radio-frequency (RF) chokes, which are corrugations in the wall that reduces radiation leakage through the cracks in the walls of the resonator~\cite{Kuo_2020NovelCavities}. As the rectangular cavity requires some method to move in the $z$-axis direction, there will be some gap between the other two axes to ensure movement. The corrugations suppress the perpendicular component of electric field at edges of the cavity, and change the mode structure such that lowest order mode becomes TM${}_{111}$-like and minimizes any incident electric field along potential gaps in the cavity.

In simulations that include it, the microstrip summing network is situated on the bottom plate of the resonator. As the microstrip is conducting metal, a trench is hollowed out to the bottom plate. A rectangular port simulating a waveguide was placed on the top of the cavity, and the second port, the readout, was situated at the CPW-SMA structure at the end of the microstrip. For simulations with a single wire antenna, the summing network is replaced with an aluminum plane without slots, and instead an antenna is fed through one of the walls.

In order to calculate the antenna coupling, the quality factor of the cavity was calculated in configurations with ($Q_{L}$) and without ($Q_{0}$) the summing network for eigenfrequency simulations of the cavity at heights from 7.1 mm to 7.9 mm. The antenna coupling is calculated from $\beta = Q_{0}/Q_{L} - 1$. For the single wire antenna, the loaded quality factor was calculated using the quality factor obtained when the antenna was modeled with some length inside. The inserted single wire antenna is placed to be parallel to the electric field of the TM${}_{110}$ mode, and centered to the shorter of the fixed rectangular cavity dimensions, similar to the antenna locations in Fig.~\ref{fig_ant_perturb}.

We will define a figure of merit that depends on the variable parameters of the cavity as follows:
\begin{equation}
    (\text{figure of merit}) = C^{2} Q_{L} \left( \frac{\beta}{1 + \beta} \right)^{2} .
\end{equation}
This figure of merit mostly focuses on $\beta$ and its possible effects on two cavity-related variables $C$ and $Q_{L}$. While the quality factor variable can be removed by using unloaded quality factor, the equation with $Q_{L}$ is used for calculations from simulation results as it is the immediate value obtainable.

\begin{figure}[h]
    \centering
    \includegraphics[width=0.9\linewidth]{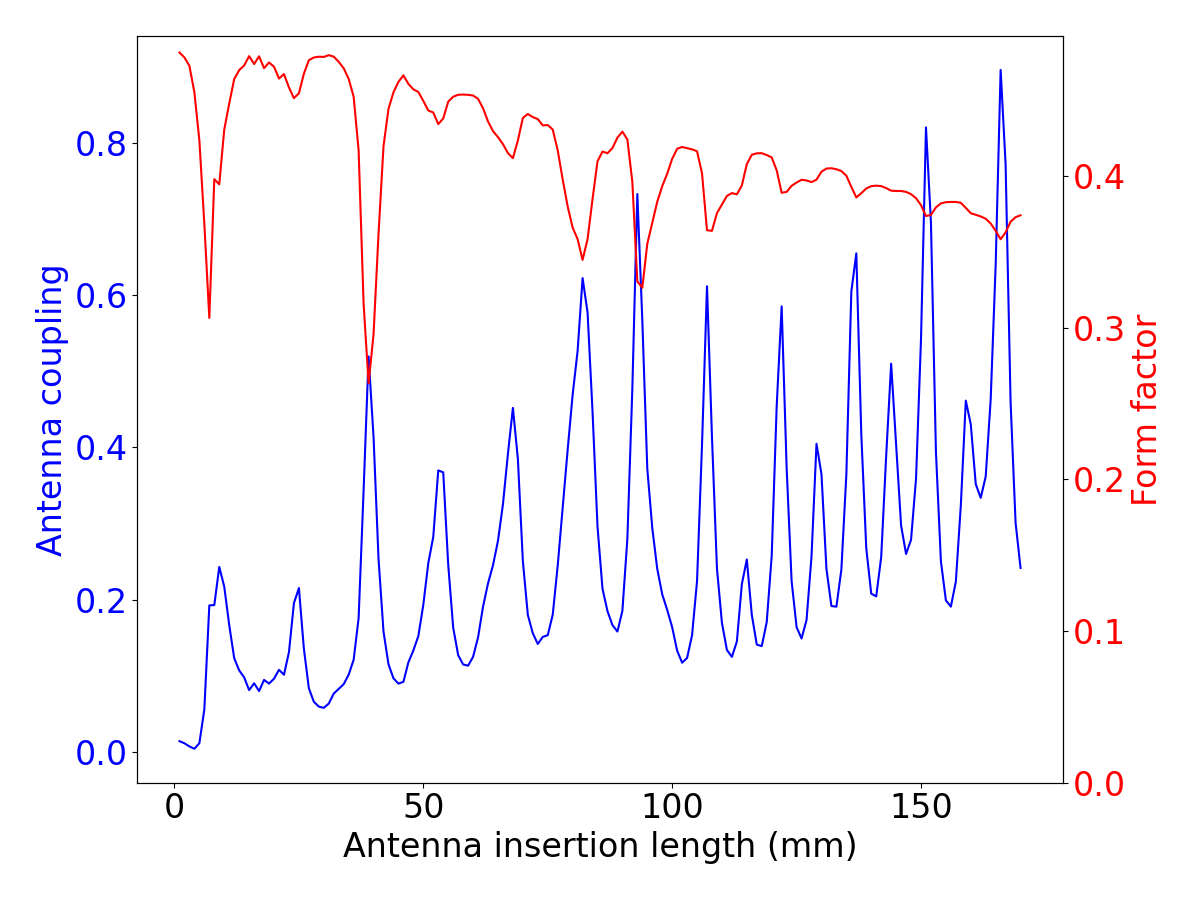}
    \caption{COMSOL simulation results for the antenna coupling (blue) and form factor (red) as a function of the length of wire inserted in the cavity. In this simulation, the height of the cavity is fixed to 7.5 mm.}
    \label{wire_coupling_single}
\end{figure}

Simulations are done for cavity heights between 7.1 mm and 7.9 mm, with a step size of 0.05 mm.  Fig.~\ref{wire_coupling_single} shows the antenna coupling and form factor of the rectangular cavity as a function of antenna length inserted into the cavity for a cavity height of 7.5 mm. In general, the antenna coupling increases with regularly spaced peaks, but does not exceed 1. The form factor decreases with antenna length, with dips at intervals that match the peaks of the antenna coupling. In consideration of both, simulation results suggest that the figure of merit is maximized towards higher insertion lengths of the wire. On the other hand, there are challenges surrounding a wire antenna inserted halfway into the rectangular cavity, in terms of being perfectly parallel to the aluminum plates without bending, and the strict requirements on insertion length to match peaks in antenna coupling. Slight deviations in both conditions affect the antenna coupling greatly, and is evident under laboratory conditions.%, which will be shown in the next section

\begin{figure}[h]
    \centering
    \includegraphics[width=0.9\linewidth]{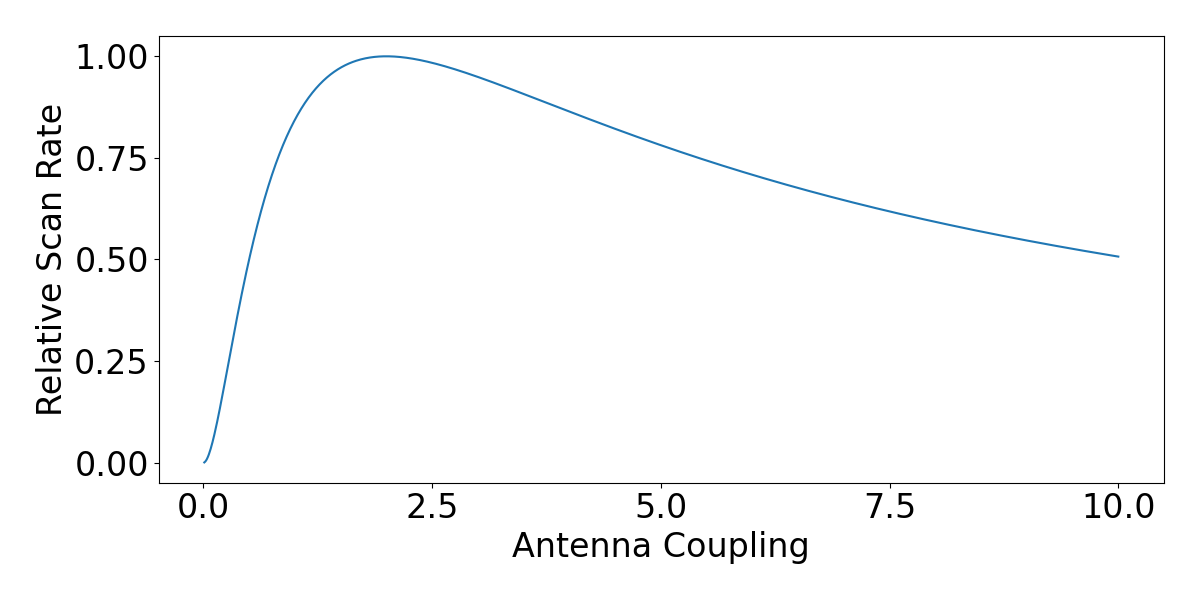}
    \caption{The scan rate as a function of antenna coupling, normalized to its maximum value when the antenna coupling $\beta$ is 2. The drop-off in scan rate at higher $\beta$ is not as steep compared to lower $\beta$ values.}
    \label{scan_rate_vs_coupling}
\end{figure}

\begin{figure} [h]
    \centering
    \includegraphics[width=0.9\linewidth]{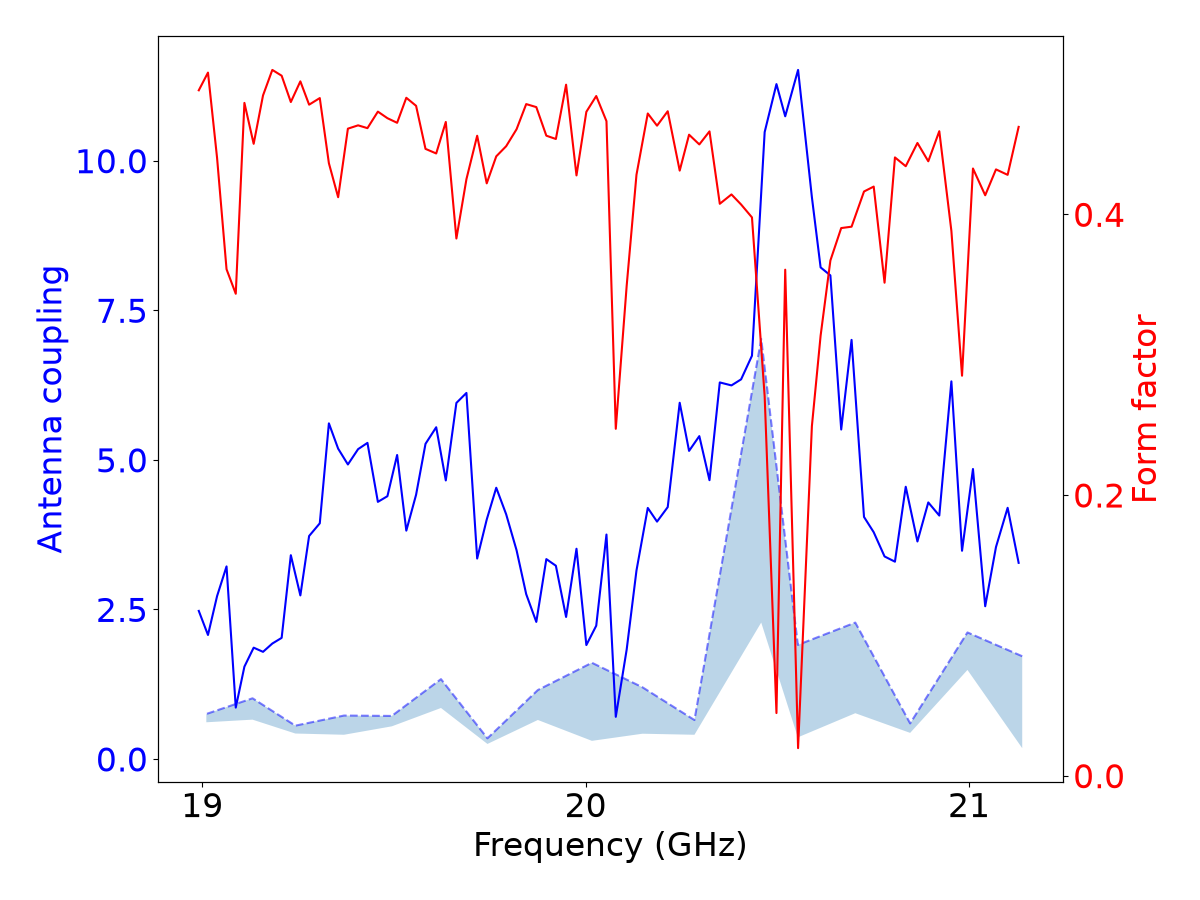}
    \caption{COMSOL simulation results for the antenna coupling (solid blue lines) and form factor (red lines) for the rectangular cavity with the microwave summing network. For comparison, a shaded blue region shows the range of antenna coupling using a single wire antenna around $\pm1$ mm of the maximum value (dashed blue lines). Simulations were done for cavity heights of 7.1 mm to 7.9 mm, and the results are plotted to their corresponding resonant frequencies of the TM${}_{111}$-like mode. Strong overcoupling of the system occurs around 20.6 GHz, which also coincides with a mode mixing (evidenced by the low form factor).}
    \label{ST_coupling}
\end{figure}

Simulations with the summing network were also done to obtain the antenna couplings for heights between 7.1 mm and 7.9 mm, with a step size of 0.01 mm. The results are shown in Fig.~\ref{ST_coupling}. As seen from the plot, the antenna coupling for the summing network is substantially overcoupled compared to the case of the single wire antenna. It is preferable that an antenna coupling surpasses 2 rather than go below for axion searches. As shown in Fig.~\ref{scan_rate_vs_coupling}, the drop-off in scan rate is slower when $\beta>2$ compared to when $\beta<2$. Moreover, at the higher frequencies of interest for this cavity, it can also be an advantage to have stronger antenna coupling at higher frequencies depending on the readout technologies used~\cite{maximizing}. As a final consideration, loss in antenna coupling due to realistic issues, such as misalignment, is expected and therefore a higher antenna coupling in the ideal case is preferred.

\begin{figure}[h]
    \centering
    \includegraphics[width=0.9\linewidth]{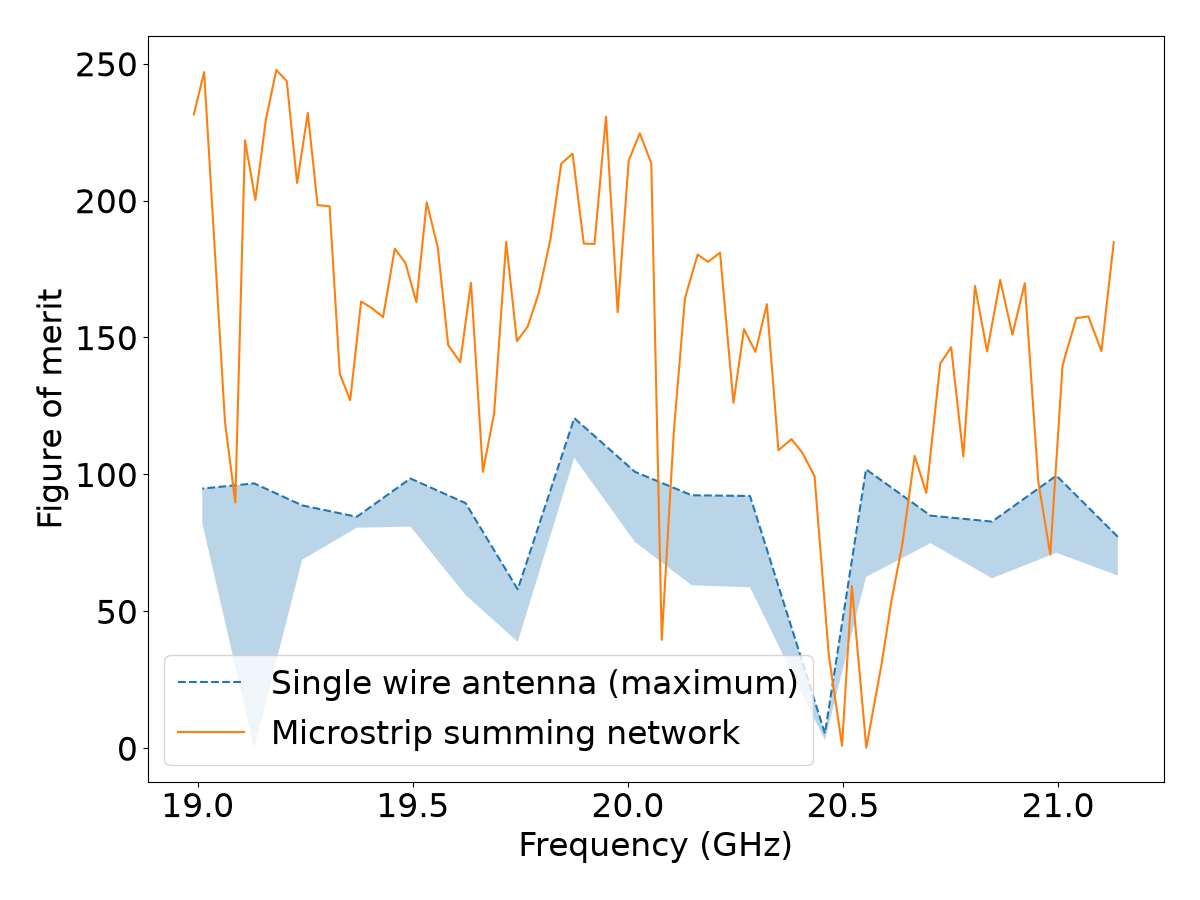}
    \caption{A comparison between figure of merits between the single wire antenna and the microstrip summing network. The shaded blue region shows the range of figure of merit using a single wire antenna around $\pm1$ mm of the maximum value (dashed blue lines). While the figure of merit between the two antenna configurations look comparable, the microstrip summing network is shows better figure of merit in general and more advantageous when subject to perturbative changes in the system.}
    \label{FOM_both}
\end{figure}

Finally, to provide a comparison between the single wire antenna and summing network, the figure of merit was calculated as a function of resonant frequency, as shown in Fig.~\ref{FOM_both}. For the single wire antenna, the maximum possible value was taken from variable antenna lengths. Despite optimizing for the best values, the figure of merit across the frequency range of $19-21$ GHz was lower compared to the microstrip summing network. 

Moreover, as mentioned above, a perturbative change to the microstrip summing network that reduces the antenna coupling would \textit{increase} the defined figure of merit since $\beta$ would be closer to 2. On the other hand, the same change would \textit{decrease} the figure of merit of a single wire antenna since the displayed values are maximized. It is also worth noting that the single wire antenna must have its length adjusted every tuning step (i.e., when the height of the cavity changes), while microstrip summing network does not require any additional adjustment. Therefore, a non-tuning summing network mitigates additional mechanical risk when coupling to the rectangular cavity, compared to the single wire antenna.

\section{Experimental Results}

In this section, we present measurement results for a 20 GHz rectangular cavity integrated with a microstrip summing network. We characterize resonant behaviors as a function of cavity tuning and with misalignment corrections. We also compare the antenna couplings achieved with this configuration to those obtained using a single-wire antenna, which is conventionally employed in lower frequency microwave cavity axion searches. All measurements are carried out at room temperature.

\subsection{Resonator Setup and Data Acquisition} \label{setup}

\begin{figure}
\centering
        \includegraphics[width=0.9\linewidth]{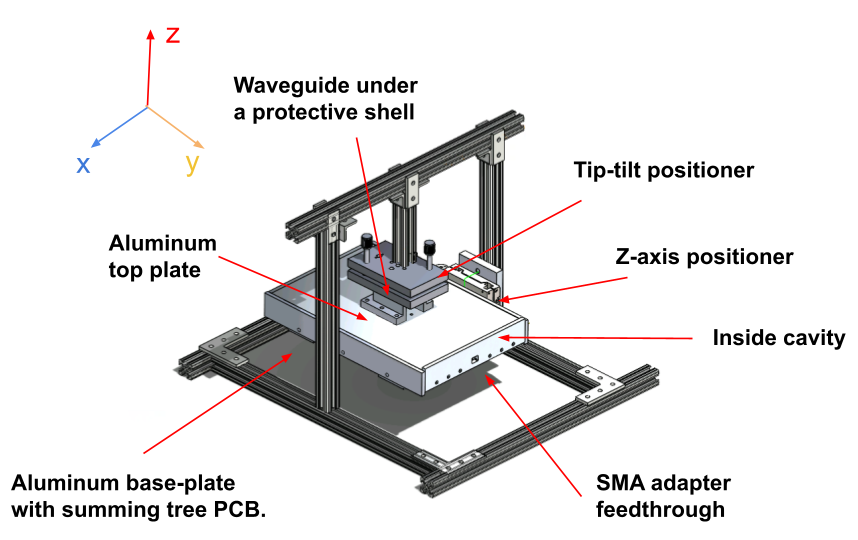}
        \includegraphics[width=0.9\linewidth]{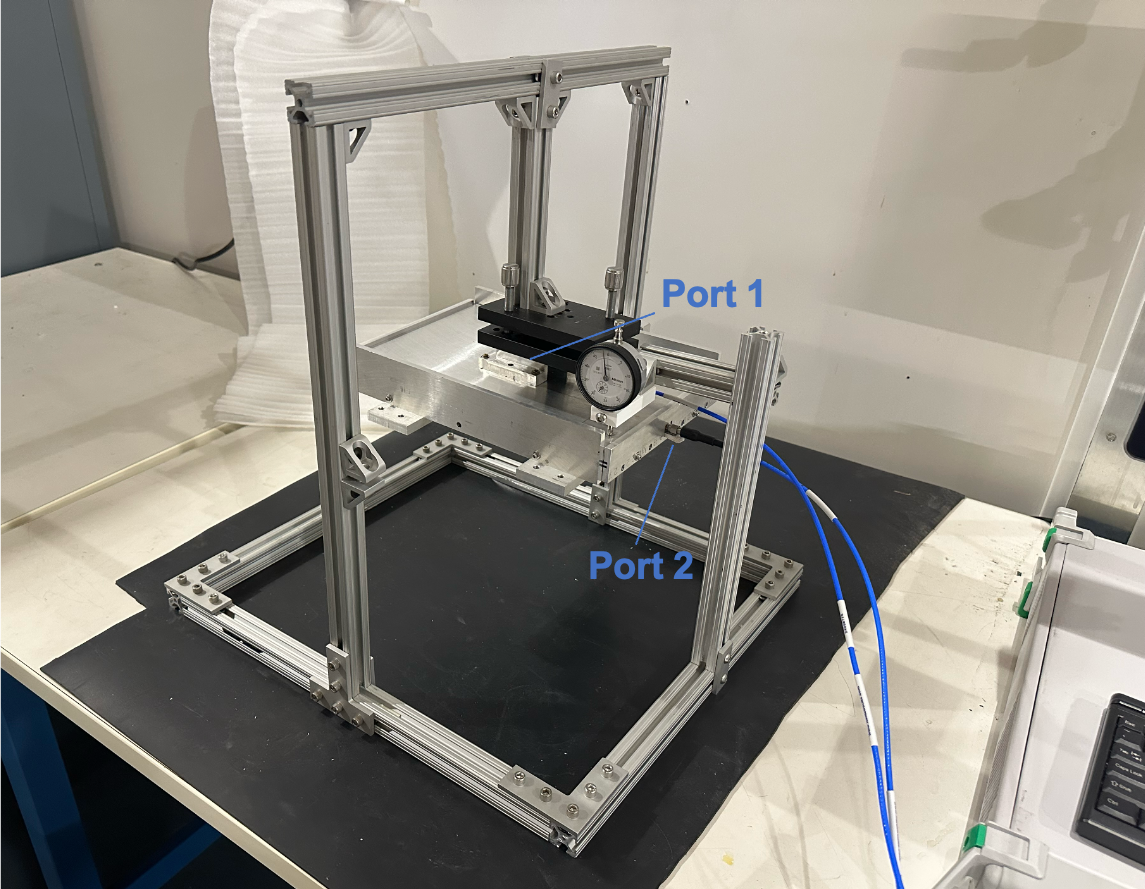}
    \caption{Top: CAD design of the resonator and 80/20 structure. Bottom: table top setup, implementing the CAD design of the resonator and the support structure.}
    \label{fig_resonator}
\end{figure}

A 20 GHz rectangular cavity integrated with the microstrip summing network was built and characterized to validate the simulation results in the previous section. Fig.~\ref{fig_resonator} shows the table-top setup. The dimensions of the cavity are manufactured to match those shown in the simulations. For tuning, we use a manual linear stage by adjusting the height ($z$-axis) of the bottom plate with an accuracy of $0.01$ mm. The resonator is also designed to angle the top plate relative to the bottom plate in order to simulate and correct the effects of misalignment between plates. For this, we use a manual tip-tilt stage and an indicator to measure the displacement at the corner of the top plate to an accuracy of $0.127$ mm. As with the $z$-axis positioner, other mechanical instabilities in the positioner result in a lower accuracy than expected from the indicator dial. Due to the mechanical imperfections and vibrations, the accuracy of our measurements are lowered in practice, but are improved by averaging multiple measurements. The whole setup is supported using an Al rail structure. This structure is placed on a rubber mat during measurements to mitigate vibrations.

The microstrip summing network design is printed on a PCB, with dimensions consistent to the values in the previous section. This was then glued on to the aluminum plate using STYCAST 2850FT. Trenches where machined into the  aluminum plate (6.35 mm deep) tracing the microstrip on the PCB, so any adhesive used during the attachment of the PCB avoids the microstrip to prevent losses. The end of the microstrip is connected to a edge launch SMA connector, and the top plate has a SMA-to-waveguide converter feeds the signal through a small circular opening on the plate, mimicking a ``weak" port in standard cavity configurations. In measurements without the summing network, the Al plate with trenches and the attached PCB is replaced with a flat Al slab. A single wire antenna is fed through the port that was originally for the summing network. 

The microwave properties of the cavity and microwave summing network readout are characterized by an Anritsu MS46322B Vector Network Analyzer (VNA). Port one of the VNA is connected to the weakly coupled port at the center of the top plate, and port two is connected to the summing network's SMA connector off the side of the cavity. This setup allows us to characterize both the summing network properties and the cavity properties by measuring the transmission from the summing network to waveguide and vice versa, and the reflection off of each port.

While $C$ can only be obtained through simulations, $Q_0$ and $\beta$ can be measured using the table-top setup. In order to quantify those values, we use a fitting function to fit the lowest order mode in the S22 data (reflection off of the summing network's SMA port) to an inverse Lorentzian:
\begin{equation} \label{inv_lorentz}
    L(\delta;A_1, A_2,\beta,Q_0)=A_1  \frac{1 - 4  \beta + A_2 Q_0 \delta}{(1 + \beta) ^2 + (Q_0\delta  \beta)^2},
\end{equation}
where $\delta=\frac{\nu}{\nu_0} - \frac{\nu_0}{\nu}$, $\nu_0$ is the central frequency of the line shape, and $A_1$ and $A_2$ are fitting parameters that correspond to the amplitude and symmetry of the line shape. This is valid for overcoupled ($\beta > 1$) reflection spectra. In the undercoupled ($\beta < 1$) case, the $\beta$ in Eq.~\ref{inv_lorentz} are replaced with $\frac{1}{\beta}$. 

We also compare standard Lorentzian fits of the S21 (transmission) data and confirmed that the loaded quality factors obtained from them agree with the values of $\beta$ and $Q_0$ from the S22 data.

\subsection{Tuning and mode identification}
\begin{figure}
    \centering
    \includegraphics[width=0.95\linewidth]{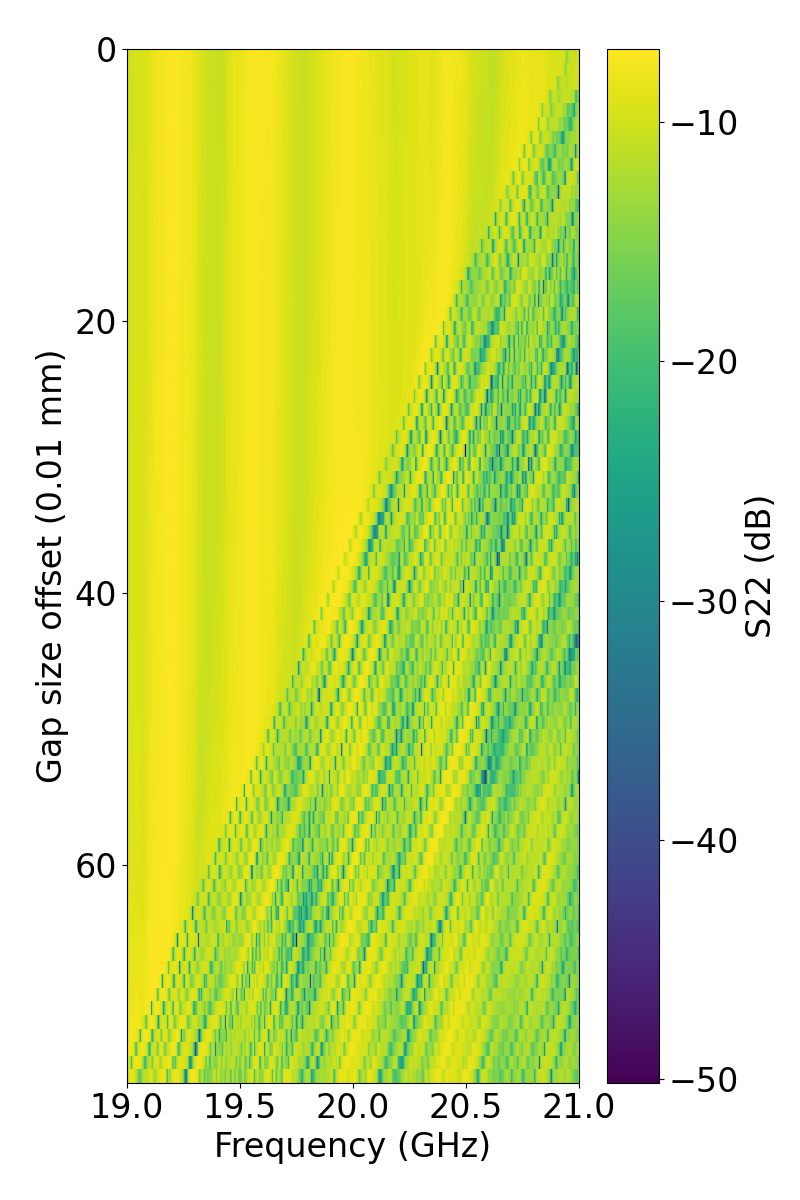}
    \caption{Mode map of $z$-axis position (the distance between the top and bottom plate) of the cavity, using reflection (S22) data. The zero value in gap size offset here refers to the starting position, not the absolute distance between the plates. The left most line is the TM${}_{111}$-like mode.}
    \label{fig_modemap}
\end{figure}

The TM${}_{110}$ mode is the lowest order resonant mode seen from the VNA, both in transmission and reflection. To locate this mode and track the change in resonant frequency as a function of height, we first create a mode map, shown in Fig.~\ref{fig_modemap}. Each step corresponds to a 0.01 mm increase in the height of the cavity. With a total of 77 steps, the frequency dependence in the table-top measurements closely matches the characteristics of the cavity in simulations, where the 2 GHz frequency tuning range was observed in cavity heights between 7.14 mm to 7.89 mm. 

Ideally, some higher order modes are also suppressed due to the symmetric nature of the slot antennas, which cancels out the signals that come from antisymmetric modes assuming good alignment.

\subsection{Alignment}
For the rectangular cavity to operate in the best experimental conditions, there needs to be a way to accurately align the top and bottom plates of the cavity. The form factor is maximized when the cavity has no misalignment, i.e. the top and bottom plates are parallel. However, since it is difficult to accurately study the electric field distribution in a realistic experimental setting, we decide to find other factors that can guide us when aligning the cavity. Specifically, we look at the resonance frequency of the lowest order mode, which will be maximized when the cavity is aligned. This assumes a pivot point for rotation about the $x$- or $y$-axis (tip or tilt) that is located at the center of the plate that is being moved.

\begin{figure}
    \centering
    \includegraphics[width=0.9\linewidth]{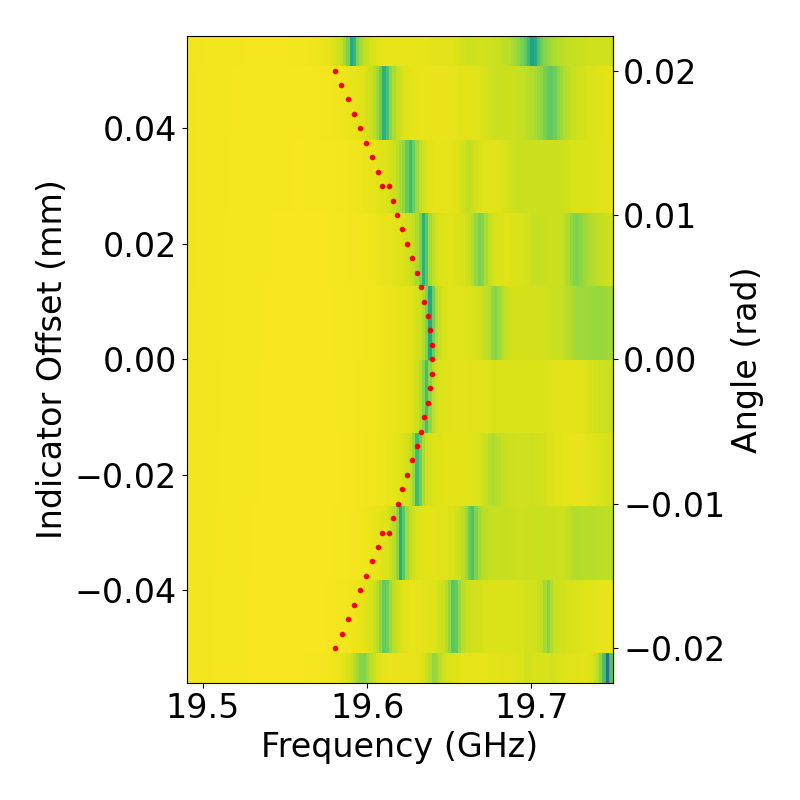}
    \includegraphics[width=0.9\linewidth]{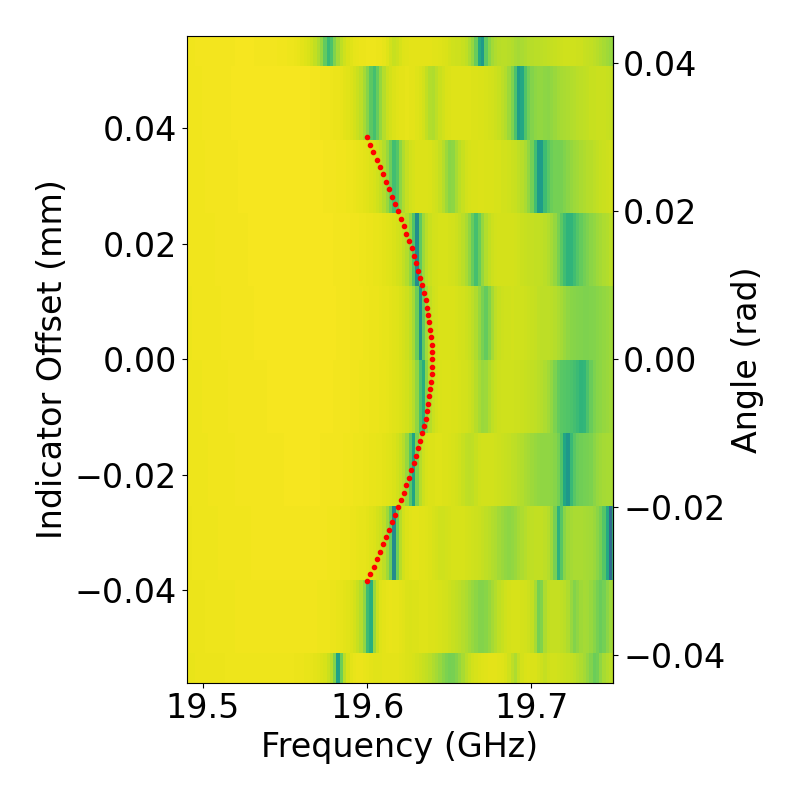}
    \caption{Top: mode map for rotation along the $y$-axis, with COMSOL simulated results overlaid for the TM${}_{111}$-like mode (dotted red line). Bottom: mode map for rotation along the $x$-axis, also with COMSOL simulated results overlaid for the TM${}_{111}$-like mode (also dotted red line).}
    \label{tiptilt} 
\end{figure}

We expect the same results in the table-top alignment. Shown in Fig.~\ref{fig_resonator}, the tip-tilt stage is attached to the top plate, and the pivot of the stage is about the center point of the plate. We compare the resonance frequency of this alignment study in Fig.~\ref{tiptilt}, which shows mode maps when there is rotation about the $x$- or $y$-axis, with COMSOL simulations overlaid (red dots) for reflection (S22) measurements.

To make these mode maps, we align the top and bottom plates near parallel, then offset them by a specified amount. We measure this displacement by using an indicator near one of the corners of the top plate, which has an accuracy of $0.0127$ mm. We then move in the opposite direction of the offset, past the parallel point, in increments of $0.0127$ mm. At small angles, a $0.0127$ mm change in the indicator position corresponds to approximately 0.01$\degree$ for rotations about the longer axis of the resonator, and 0.005$\degree$ for rotations about the shorter axis of the resonator.

The COMSOL results show good qualitative agreements with the measurements and the remaining differences are likely due to the mechanical instability in the table-top setup. Although limited somewhat by the simple support structure and the manual adjustments, these results verify that the fundamental mode is clean, tunable, and that an 8-slot summing network can couple more strongly to the cavity modes than a simple wire antenna.  These measurements also demonstrate the importance of motorized alignment and vibration isolation in future implementations of axion searches.  
For example, using the property that the resonance frequency is maximized when the two plates of the cavity are parallel, alignment can be guided by an automated Nelder-Mead algorithm, similar to the optimization for multidimensional optimization used with the single wedge cavity in Ref.~\cite{singleWedge_Taj2024}.

\subsection{Coupling $\beta$ and Quality Factor $Q$}
\label{sec_beta_and_Q}

We proceed to measure the antenna coupling and quality factor the tuning range of interest for the cavity.  

\begin{figure}[h]
    \centering    \includegraphics[width=0.9\linewidth]{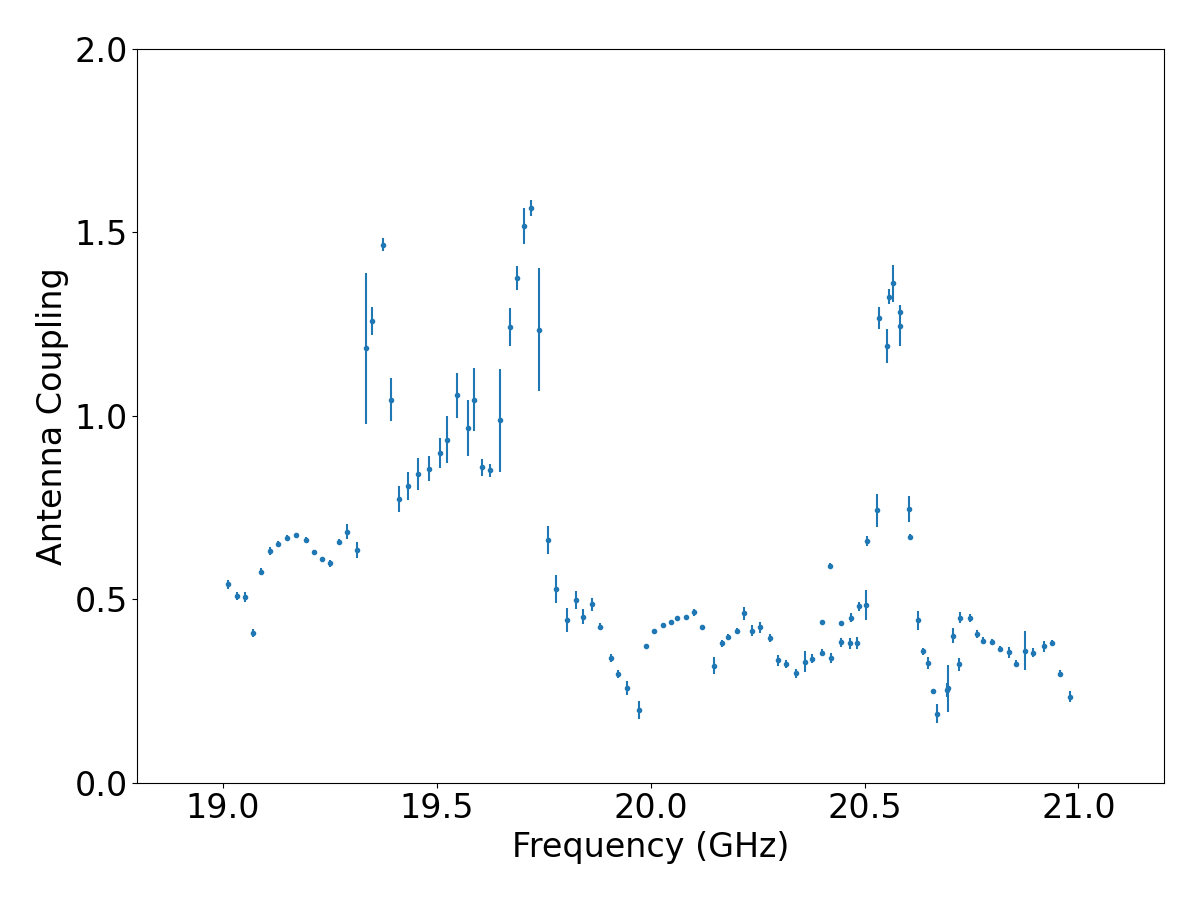}
    \includegraphics[width=0.9\linewidth]{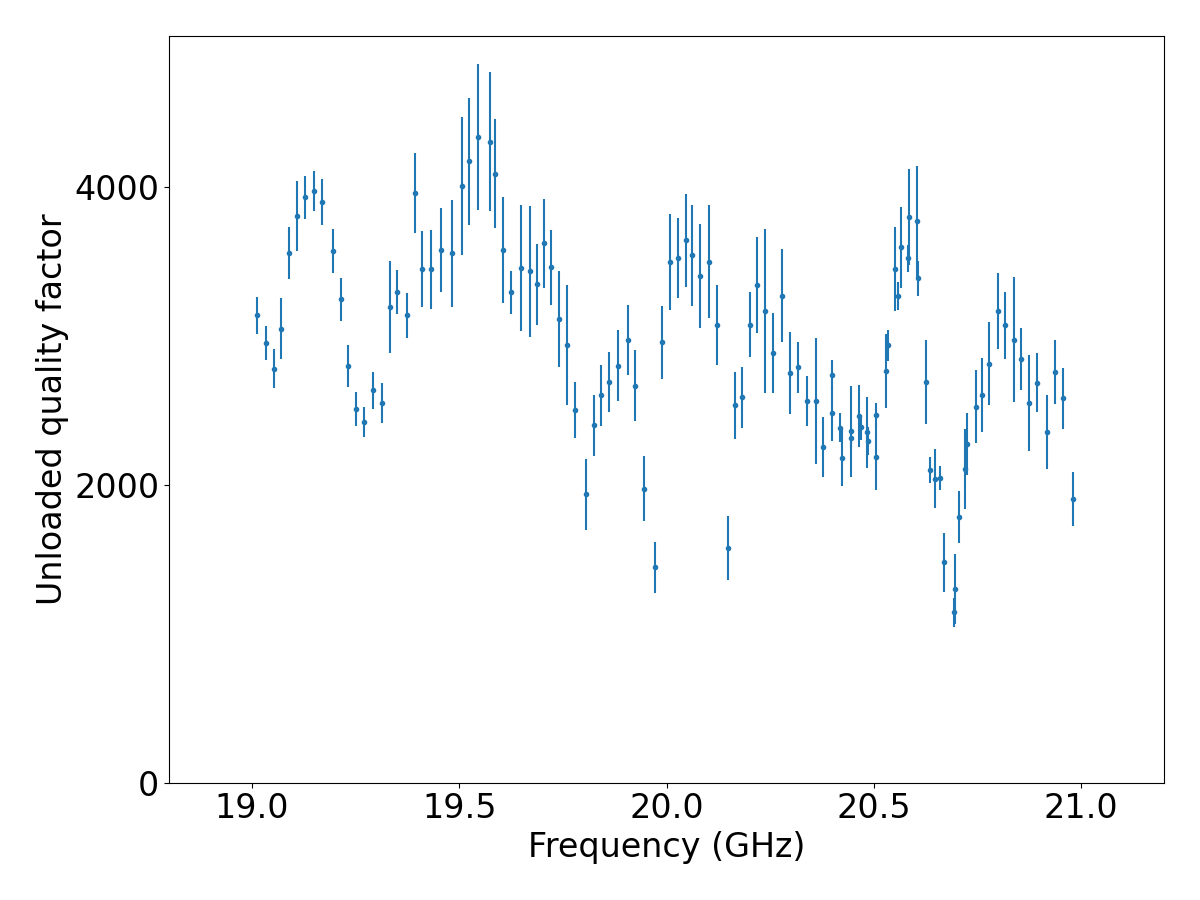}
    \caption{Top: Measured antenna coupling $\beta$ across the whole $19 - 21$ GHz range. Bottom: Measured unloaded quality factor $Q_{0}$ across the same frequency range. As there were multiple repeated measurements at various tuning steps, there are error bars in the measurements for each frequency bin of interest.
    }
    \label{fullSweep}
\end{figure}

Fig.~\ref{fullSweep} shows the measured $\beta$ and derived $Q_0$. We see that, in general, $\beta>0.3$, with increases at frequencies that are in qualitative agreement with the simulated values in Fig.~\ref{ST_coupling}. We can also see from this figure that $Q_0$ is on average about 3000, except for a few values near 20 GHz and 20.6 GHz.

Comparing the measured $Q_0$ and $\beta$ to the simulations, we see that the achieved results here are generally lower. The summing network has an antenna coupling $\beta$ that is between 2 and 6 in simulations and between 0.3 and 2 in measurements. On the other hand, the single wire antenna has a coupling that is 0.2 or lower unless its probe length is tuned to very specific values.
In practice, the rectangular cavity was also measured using a single wire antenna, but the lowest order mode did not show any noticeable reflection, showing that the coupling was negligible. The additional loss in single wire antenna coupling was due to the fact that the antenna wire length was fixed, making matching the single wire probe to its best couplings extremely difficult, and the non-straightness of the probe.

Recalling the scan rate's dependence on antenna coupling in Eq.~\ref{eq_scan_rate_beta}, the 8-slot summing network offers a significant upgrade over a typical wire antenna for microwave cavity readout, even accounting for losses due to an imperfect experimental setup. Possible future upgrades and prospects of using the summing network as part of an axion search will be discussed in Sec.~\ref{sec_discussion}.

\section{Discussion} \label{sec_discussion}

From the numerical studies and the subsequent experimental confirmation, we show that a microstrip summing network can achieve a greater $\beta$ compared to a single-wire antenna, which mostly stays in the undercoupled regime in all configurations. In the current implementation, the antenna coupling values of the 8-slot microstrip summing network are still suboptimal. With better geometrical tolerances and additional slots in the summing network, we can expect to increase $\beta$ even further and approach an optimal coupling. 

Additionally, axion detectors are typically operated at cryogenic temperatures in order to reach the KSVZ or DFSZ limits. Ultimately, cavities like the wedge cavity~\cite{singleWedge_Taj2024}, for which the summing network could be used as the readout system, will be cooled by a dilution fridge in a cryostat. PCBs are known to be cryogenically compatible. In regard to the shortcomings of the prototype cavity with the summing network, a higher quality factor can be achieved through the use of more conductive metals and operation at lower temperatures.

On the other hand, many of the mechanical challenges faced in this work would be different under such operating conditions. The current summing network design is also prone to mechanical stress and other issues that can lead to additional loss. Some of the losses in antenna coupling value could be recovered with a setup having tighter mechanical precision. Alternative designs are being made that utilize waveguides, rather than transmission lines, to reduce these stresses and losses. All of the above improvements will lead to a more consistent targeted antenna coupling value.

\begin{figure*}
    \centering
    \begin{minipage}{.9\textwidth} 
        \centering
        \includegraphics[width=0.9\linewidth]{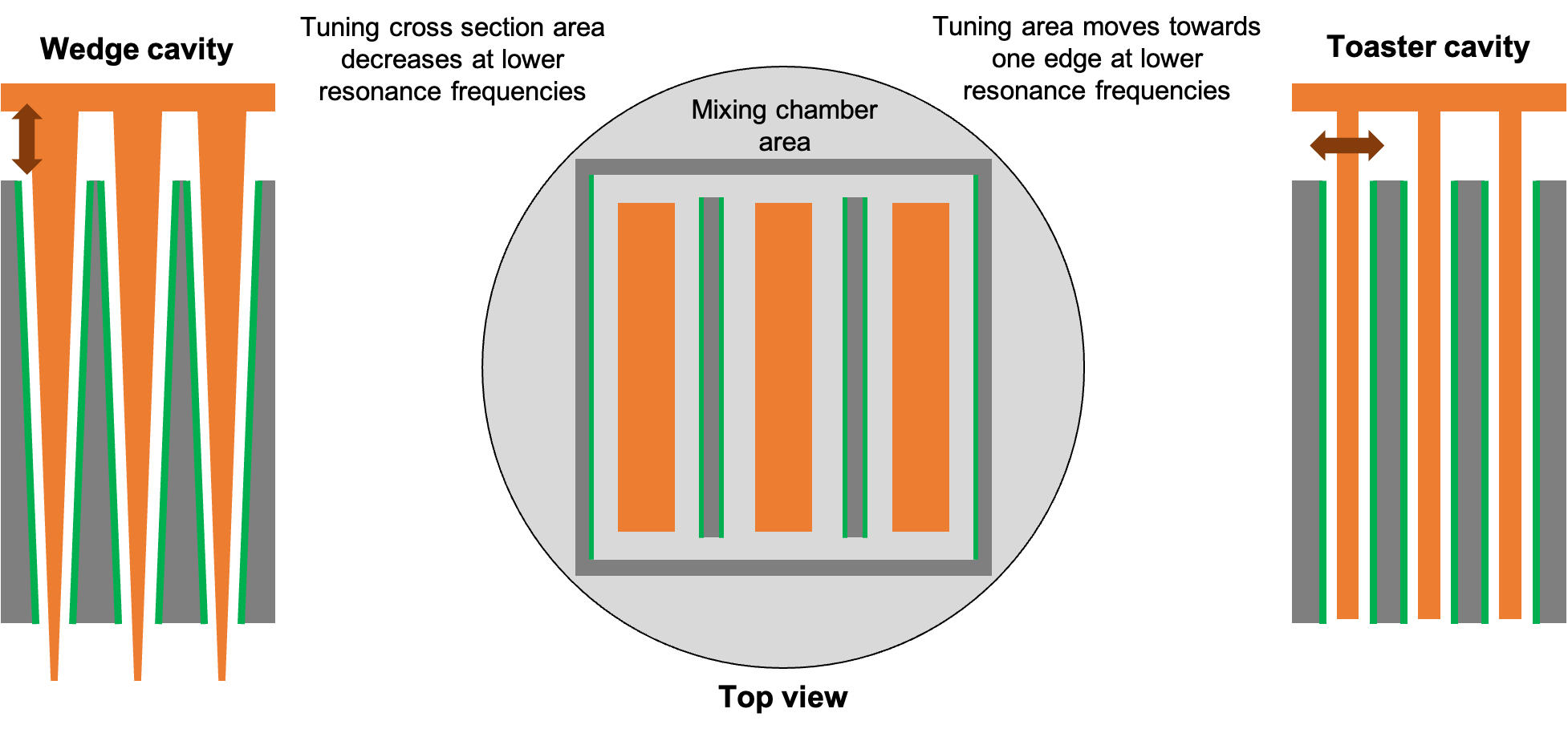}
    \end{minipage}
    \hfill 
    \caption{A schematic of two possible ADMX-VERA configurations which could utilize a microstrip summing network readout. Both configurations, named the wedge cavity (left) and toaster cavity (right), respectively, have fixed cavity and inner walls, shown in dark gray. Lines shown in green depict the placement of summing networks, which would coherently combine into a signal. There is a tuning element, shown in orange, which moves to change the resonance frequency of the system. Both configurations are not shown to scale for a 20 GHz cavity, as this would have more volumes (i.e., more wedges or toaster ``slots.")}
    \label{fig_schematic}
\end{figure*}

\begin{figure}[ht!]
\centering
    \includegraphics[width=0.9\linewidth]{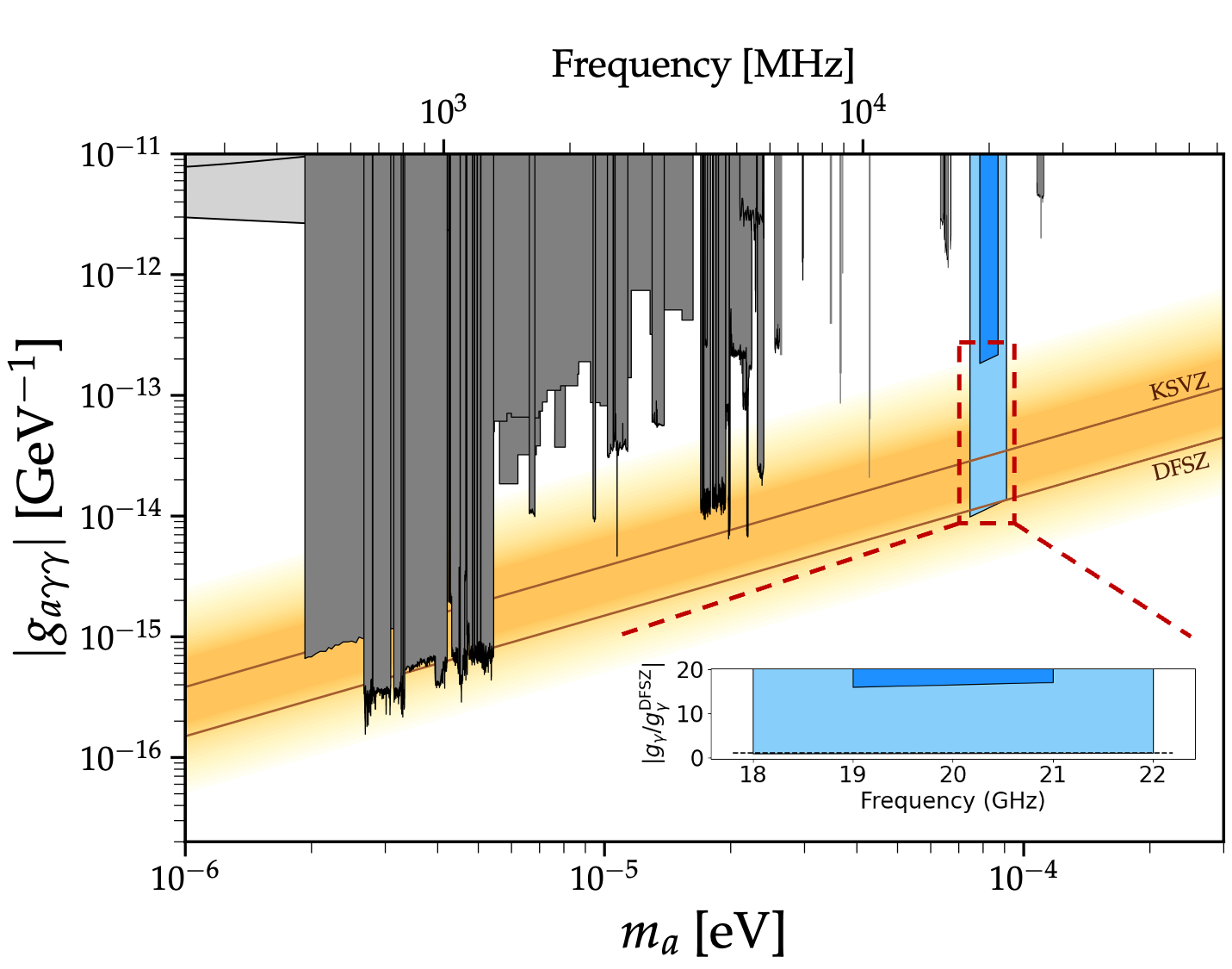}
    \caption{Discovery potential of the microstrip summing network. The sensitivity is shown in comparison to the benchmark KSVZ model. Shown in blue is a cryogenic version of the currently shown microstrip network, with the parameters discussed in the main text. The projection shown in light blue utilizes a multi-wedge cavity with volume dimensions that aim to maximize a 0.4 m diameter volume. The inset shows the projections as a closeup around 20 GHz in multiples of DFSZ given the parameters shown in the text. The plot was generated by an adapted version of the
    software provided by~\cite{ohare_software}.}
    \label{fig_exclusion}
\end{figure}

With all of the above considerations in mind, Fig.~\ref{fig_schematic} shows the possible configurations of a cavity that uses the microstrip summing network, and Fig.~\ref{fig_exclusion} shows its discovery potential. For the science reach of an axion search, we initially consider the possible sensitivity reach of a single rectangular cavity with the same dimensions as this work in an environment similar to that of ADMX. The system assumes a 7.6 T magnet and a cryogenic system that cools down to 100 mK with added noise at the standard quantum limit (SQL). The cavity has an unloaded quality factor $Q_{0}\sim16500$ following the scaling set by Ref.~\cite{maximizing}, a form factor $C=0.4$, and the output assumes a summing network that maintains antenna coupling at $\beta=2$. As seen in the simulations, there are regions where the antenna coupling may be stronger, but as this can be changed by the number of slots or their distribution. Following the comparison standards set by Ref.~\cite{maximizing}, the data taking time is paced to be 100 days over the 5\% fractional bandwidth. Since the tuning range of the cavity in this work actually has a 10\% bandwidth, the total data taking time amounts to 200 days. The signal-to-noise ratio for signal detection is set to 5.

Evidently, a single rectangular cavity does not utilize the full volume (about 0.4 m in diameter) of the magnetic system used for ADMX axion searches. In order to fill the volume more efficiently, we will require multiple volumes of a cavity resonant at 20 GHz, especially since the width of the cavity volume is constrained to be around 7.5 mm for this geometry. The aforementioned wedge cavity is one feasible example. By applying the slot antennas to the fixed walls of the cavity, it would be possible to increase the antenna coupling proportional to the volume. Similarly, a configuration where tuning slabs move horizontally closer or further away from the cavity walls (named the ``toaster" cavity) can also produce similar resonance frequencies in an expanded volume. In this configuration, a volume is partitioned with flat, rectangular walls, and there are rectangular slabs that act as ``tuning rods" in between the walls. By moving the slabs closer towards one wall, the resonant frequency of the volume decreases, as most of the electric field is pushed toward the side where the distance between the wall and slab is wider. The form factor of the cavity will suffer in the middle of its tuning range, when the distances between the cavity walls and tuning slabs are different on each sides. However this geometry might be able to utilize the $\sim 1$~m tall vertical space more efficiently than the wedge cavities.

In the proposed magnet-filling ADMX-VERA axion search, we assume that an ideal wedge or toaster cavity has the same quality factor and form factor. The gain in volume is about 30 times that of the single rectangular cavity. 
We further assume the use of beyond-SQL technologies for readout, e.g., taking advantage of the scan rate enhancement provided by single microwave photon detectors (SMPD). As seen in Ref.~{\cite{maximizing}, a higher antenna coupling from the cavity is preferred, which is significantly easier to achieve through the placement of multiple slots inside this cavity as compared to a single wire antenna. Assuming a dark count rate of 100 per second and setting the antenna coupling $\beta$ to 10, the scan rate increases by a factor of 92. Through all these enhancements, the DFSZ model axions would become accessible in the 20 GHz frequency band. 

\section{Conclusion}
Using realistic numerical simulations and bench top demonstrations, we show that a microstrip summing network enhances coupling to a high frequency ($\sim$ 20 GHz) $V\gg \lambda^3$ resonant cavity. The 8-slot tree studied is able to maintain a good coupling with moderate misalignments within the cavity. The measured coupling $\beta$, although lower than what the simulations predicts, is already a significant improvement over the wire probe coupling. The improvement in $\beta$ is very promising both for generally coupling to the cavity, and being able to couple to localized modes. 

Further development in the summing network will involve adding additional slots, such as a 96 slot array, as well as trying different transmission methods such as waveguides, rather than micro strips on a PCB. Additional antennas can be used to increase the coupling, and therefore we are confident that this is a viable alternative to the single wire coupling. By replacing the transmission line with a waveguide, we could also reduce the attenuation throughout the summing network.

\section{Acknowledgments}
The paper is based upon work supported by the National Science Foundation under Grants No. 2209576 and 2513664. Chelsea Bartram, Andrew Yi, Jacob Laurel, and the detector and data acquisition components listed in this manusript were supported by the Department of Energy, Laboratory Directed Research and Development program at SLAC National Accelerator Laboratory, under contract DE-AC02-76SF00515 and as part of the Panofsky Fellowship awarded to Chelsea Bartram.

\bibliographystyle{apsrev4-1}
\raggedright
\bibliography{references}

\end{document}